\documentclass{article}
\usepackage{graphicx,epsfig,rotating,amssymb}

\newlength{\figwidth}
\newcommand{\dfrac}{\displaystyle \frac}
\newcommand\GeV{\ifmmode {\mathrm{\ Ge\kern -0.1em V}}\else
                   \textrm{Ge\kern -0.1em V}\fi}
\begin{document}

\setlength{\unitlength}{1mm} \thispagestyle{empty}


\vspace{2cm}

\begin{center}
{\Large \textbf{Absolute determination of $D_s$ branching ratios and $f_{D_s}$ extraction at a neutrino factory}}
\end{center}

\vspace{2cm}


\begin{center}
\textbf{\large G. De Lellis$^{1)}$, P. Migliozzi$^{2)}$ and P. Zucchelli$^{3)}$}\\
\vspace{0.2cm} {1) Universit\`{a} Federico II and INFN, Napoli, Italy}\\
{2) INFN, Napoli, Italy}\\
{3) CERN, Geneva, Switzerland and INFN, Ferrara, Italy}

\begin{abstract}
A method for a direct measurement of the exclusive $D_s$
branching ratios and of the decay constant $f_{D_s}$ with a
systematical error better than $5\%$ is presented. The approach is
based on the peculiar vertex topology of the anti-neutrino induced
diffractive charm events. The statistical accuracy achievable with a
neutrino factory is estimated.
\end{abstract}
%
%
\end{center}

\newpage
%
%

\section{Introduction}

The experimental knowledge on leptonic $D_s$ decays is rather 
poor. Currently, the branching ratios for $D_s\rightarrow l\nu$
decays are estimated by the Particle Data Group~\cite{pdg2000} to be
$BR(D_s\rightarrow \mu\nu) = (4.6\pm1.9)\times10^{-3}$ and
$BR(D_s\rightarrow \tau\nu) = (7\pm4)\times10^{-2}$. These large
uncertainties translate into a large uncertainty on the extraction of
the decay constant $f_{D_s}$.

In this paper we propose a method to build an almost pure sample of
 $D^-_s$ from diffractive events which allows the extraction of most
 of the $D_s$ branching ratios and in particular of purely leptonic
 decays from which $f_{D_s}$ can be extracted with a systematic error
 better than $5\%$.  A statistics capable of exploiting this
 systematic error can be accumulated at neutrino factories. Once
 $f_{D_s}$ is measured with such an accuracy, one would feel more
 confident about extrapolating to the decay constants in the $B$
 system, $f_{B}$ and $f_{B_s}$, which are crucial quantities for a
 quantitative understanding of $B^0_{(s)} - \bar{B}^0_{(s)}$
 oscillations and the possible extraction of the CKM matrix elements
 $V_{td}$ or $V_{ts}$~\cite{bigi}.

This paper is organised as follows: in Section~\ref{diff_pro} we
discuss the available data on neutrino and anti-neutrino induced
diffractive $D_s^{(*)}$
\footnote{In the following $D_s^{(*)}$ stands for either $D_s$ or $D_s^{*}$. 
The same notation is also used for $D^{(*)}$.} production. In
Section~\ref{leptodec} the leptonic $D_s$ decay mechanism is discussed
and the available experimental determinations of $f_{D_s}$ are
reviewed.  A method for a direct evaluation of
$D_s$ branching ratios and $f_{D_s}$ measurement is discussed in
Section~\ref{accu}, together with the evaluation of the accuracy
achievable at a neutrino factory.  In the last section we give our
conclusions.

%
%
\section{ (Anti-)Neutrino induced diffractive $D_s^{(*)}$ production}
\label{diff_pro}

\begin{figure}[htbp]
\begin{center}
\rotatebox{0}{\ \resizebox{0.4\textwidth}{!}{\ \includegraphics{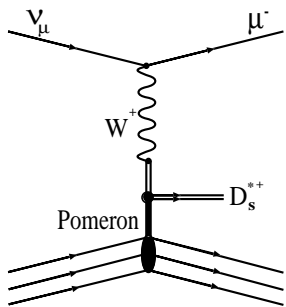} }}
\end{center}
\caption{Diagram for neutrino induced diffractive $D_s^{*+}$ meson production.}
\label{fi:ds_star}
\end{figure}

In charged-current (CC) interactions, the $W$ boson can {\em
fluctuate} into a charmed meson. The on-shell meson is produced by
scattering off a nucleon or a nucleus without breaking up the
recoiling partner. The diffractive production mechanism ($\nu_\mu N
\rightarrow \mu^-D_s^{*+} N$) is shown schematically in
Fig.~\ref{fi:ds_star}. The same mechanism applies to $D^{(*)}$
production, but the $D_s^{(*)}$ one is Cabibbo favoured by a
factor $\left |\frac{V_{cs}}{V_{cd}}\right|^2\sim20$.

This process has been observed in previous
experiments~\cite{dsprod,dsprod_nutev,chorus_ds}. In particular, 
one of these experiments~\cite{chorus_ds} has shown the evidence of a 
$D_s$ diffractive production through the direct observation in nuclear 
emulsion of the decay chain $D_s^{*+} \rightarrow D_s^{+} \gamma $, 
$D_s^{+} \rightarrow \tau^+ \nu_{\tau} $, $\tau^{+} \rightarrow \mu^{+} 
\bar{\nu}_{\tau} \nu_{\mu}$. 

Unlike deep-inelastic cross-sections, the diffractive cross-sections
are predicted to be the same for $\nu$ and $\bar{\nu}$ and for proton
and neutron targets. The observed neutrino and anti-neutrino induced
diffractive $D_s$ rates relative to CC interactions on an isoscalar
target have been measured to be
$(1.5\pm0.5)\times10^{-4}~\mbox{and}~(2.6\pm0.9)\times10^{-4}\mbox{,
respectively}$. This is compatible with a $1:2$ ratio, as implied by
the equality of diffractive cross-sections and the ratio of inclusive
cross-sections~\cite{dsprod}. It is worth stressing that these numbers
have been obtained searching for particular $D_s$ decay channels in
$D_s^{(*)}$ diffractive production: therefore the corresponding
branching ratio has to be known to get the absolute production
rates~\cite{dsprod}.

The energy dependence of the diffractive cross-section has also been 
investigated. In the neutrino energy intervals $10\div30$, $30\div50$
and $50\div200$~GeV the observed rate per CC is
$(1.8\pm0.7)\times10^{-4}$, $(1.3\pm0.6)\times10^{-4}$ and
$(1.6\pm0.7)\times10^{-4}$, respectively, so that no variation is
detected at the available statistical
level~\cite{dsprod}. Theoretically, no steep variation of this
relative rate in the $10\div200$~GeV neutrino energy interval is
expected.

Taking into account the corresponding branching ratio, the neutrino
diffractive $D_s^{(*)}$ production rate has been evaluated to be
$(2.8\pm1.1)\times10^{-3}$/CC~\cite{dsprod}.  An independent
evaluation of the same production rate has given
$(3.2\pm0.6)\times10^{-3}$/CC~\cite{dsprod_nutev}.  The weighted
average of the two evaluations gives a neutrino induced diffractive
production rate of $(3.1\pm0.5)\times10^{-3}$/CC.  Since the neutrino
to anti-neutrino CC production rate is $2:1$, the average
anti-neutrino induced $D_s^{(*)}$ diffractive production rate is
$(6.2\pm1.0)\times 10^{-3}$/CC. Therefore, the combined analysis gives
an accuracy of about $15\%$ for the diffractive production rate.

For a detailed theoretical review of lepton induced diffractive
production we refer to Refs.~\cite{teods1,teods2,teods3,teods4,teods5}.

%
%
\section{Leptonic $D_s^+$ decays}
\label{leptodec}
From a theoretical point of view, purely leptonic decays of charged
mesons are the simplest ones to describe.  The effect of the strong
interaction can be parameterised in terms of just one factor, called
the decay constant. Unlike semi-leptonic decays, where $q^2$ (and
hence the form factors) varies event by event, leptonic decays
have a fixed $q^2$ value ($q^2=M^2$, where $M$ is the mass of the
initial meson).

\begin{figure}[htbp]
\begin{center}
\rotatebox{0}{\ \resizebox{0.5\textwidth}{!}{\ \includegraphics{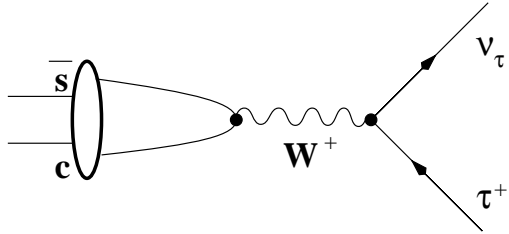}}}
\end{center}
\caption{Feynman diagram of the $D_s^{+}\rightarrow\tau^{+} \nu_{\tau}$ 
decay.}
\label{fi:dsdecay}
\end{figure}

By neglecting radiative corrections, the decay rate of a charged
pseudo-scalar meson $M_{Q\bar{q}}$ to $l {\nu_{l}}$ is

$$\Gamma(M_{Q\bar{q}} \rightarrow l \nu_{l}) =
\frac{G_F^2}{8\pi}\times\mid V_{Qq}\mid^2\times f_M^2\times M
\times m^2_l\times\left(1-\frac{m_l^2}{M^2} \right)^2$$

where $f_M$ is the decay constant, $ V_{Qq}$ is the CKM matrix
element, and $m_l$ and $M$ are the masses of the lepton and charged
meson $M_{Q\bar{q}}$, respectively. The decay constant $f_M$ is a
measurement of the probability amplitude for the quarks to have zero
separation, which is necessary for them to annihilate. 

By using the previous notation, the $D_s$ leptonic branching ratio can
be written as

$$BR(D_s \rightarrow l^-\bar{\nu}_l) = \frac{G_F^2}{8\pi}\times\mid V_{Qq}\mid^2\times f_{D_s}^2\times \tau_{D_s}\times M_{D_s}\times m^2_l\times\left(1-\frac{m_l^2}{M_{D_s}^2} \right)^2$$

where $ \tau_{D_s}$ is the $D_s$ life-time.

Decay constants for pseudo-scalar mesons containing a heavy quark have
been predicted with lattice QCD, QCD sum rules and potential models,
but due to the non-perturbative character of the calculations they vary 
significantly in
the predictions. The predicted value for $f_{D_s}$ lies in the 
$190\div360$~MeV range~\cite{richman}.

A method for extracting $f_{D_s}$ is to measure the leptonic decay
modes of $D_s$. Due to helicity suppression, only the muonic and
tauonic decay modes have an appreciable branching ratio. Unfortunately,
the measurements of the leptonic $D_s$ decays are rather scarce. A
summary of the available measurements together with the determinations
of $f_{D_s}$ is shown in Table~\ref{tab:fds}. The errors on $f_{D_s}$ are
larger than $30\%$ and, despite the scanty statistics available, the
systematic error dominates. This is mainly due to the uncertainty on the
normalisation used for the determination of the leptonic decay
branching ratios. 

Once the absolute value of $f_{D_s}$ is measured with
an accuracy of $10\%$ or better, one would feel more confident about
the predictions of the decay constants in the $B$ system, $f_{B}$ and
$f_{B_s}$, which are crucial quantities for a quantitative
understanding of $\bar{B}^0_{(s)}-B^0_{(s)}$ oscillations and the extraction of
$V_{td}(V_{ts})$ from them.

\begin{table}[tbp]
\begin{center}
{\small
\begin{tabular}{||c|c|c||}
\hline
Experiment & Channel & $f_{D_s}~(\mbox{MeV})$ \\ \hline\hline
WA75\cite{wa75} & $D_s\rightarrow\mu$ & $232\pm45\pm52$ \\ \hline
CLEO I \cite{cleo1ds} & $D_s\rightarrow\mu$ & $344\pm37\pm52\pm42$ \\ \hline
CLEO II\cite{cleo2ds} & $D_s\rightarrow\mu$ & $280\pm19\pm28\pm34$ \\ \hline
E653\cite{e653} & $D_s\rightarrow\mu$ & $194\pm35\pm20\pm14$ \\ \hline
BEATRICE\cite{beatrice} & $D_s\rightarrow\mu$ & $323\pm44\pm12\pm34$ \\ \hline
BES\cite{bes} & $D_s\rightarrow l$ & $430^{+150}_{-130}\pm40$ \\ \hline
L3\cite{l3} & $D_s\rightarrow\tau$ & $309\pm58\pm33\pm38$ \\ \hline
\end{tabular}
}
\end{center}
\caption{Summary of the available experimental determinations of the decay constant  $f_{D_s}$.}
\label{tab:fds}
\end{table}

In the following we describe a method for the extraction of the $D_s$ 
leptonic branching ratios  which minimises the systematical error due to 
the normalisation of the sample. It is based on the direct observation 
of $D_s^{(*)}$ produced in anti-neutrino induced diffractive interactions 
and a topological selection of the events which improves the purity of 
the sample used for the normalisation. 

%
%
\section{Direct evaluation of $D_s$ branching ratios and $f_{D_s}$ measurement}
\label{accu}
In the following we consider a detector with the capability to exploit 
vertex and decay topologies with micron  precision. We assume to use 
an emulsion  target (see for instance~\cite{chorus}), 
although the method we are proposing would
  work with a CCD target~\cite{bigi} as well. This allows the detection of 
  short lived particles with path length larger than $10~\mu m$. In order to 
measure the charge and the momentum of hadrons and muons, electromagnetic and 
hadronic showers, a possible detector is the one described in Ref.~\cite{bigi} 
once equipped with such an emulsion target.

\subsection{Event yield and flux at a neutrino factory}

For this study we assume a muon beam energy $E_\mu = 50$~GeV; length 
of the straight section, $L = 100$~m; muon beam angular divergence, 
$0.1\times m_\mu/E_\mu$, $m_\mu$ being the muon mass; muon beam transverse 
size $\sigma_x = \sigma_y = 1.2$~mm. 

\begin{figure}[tb]
\begin{center}
\rotatebox{0}{\ \resizebox{1.0\textwidth}{!}{\
\includegraphics{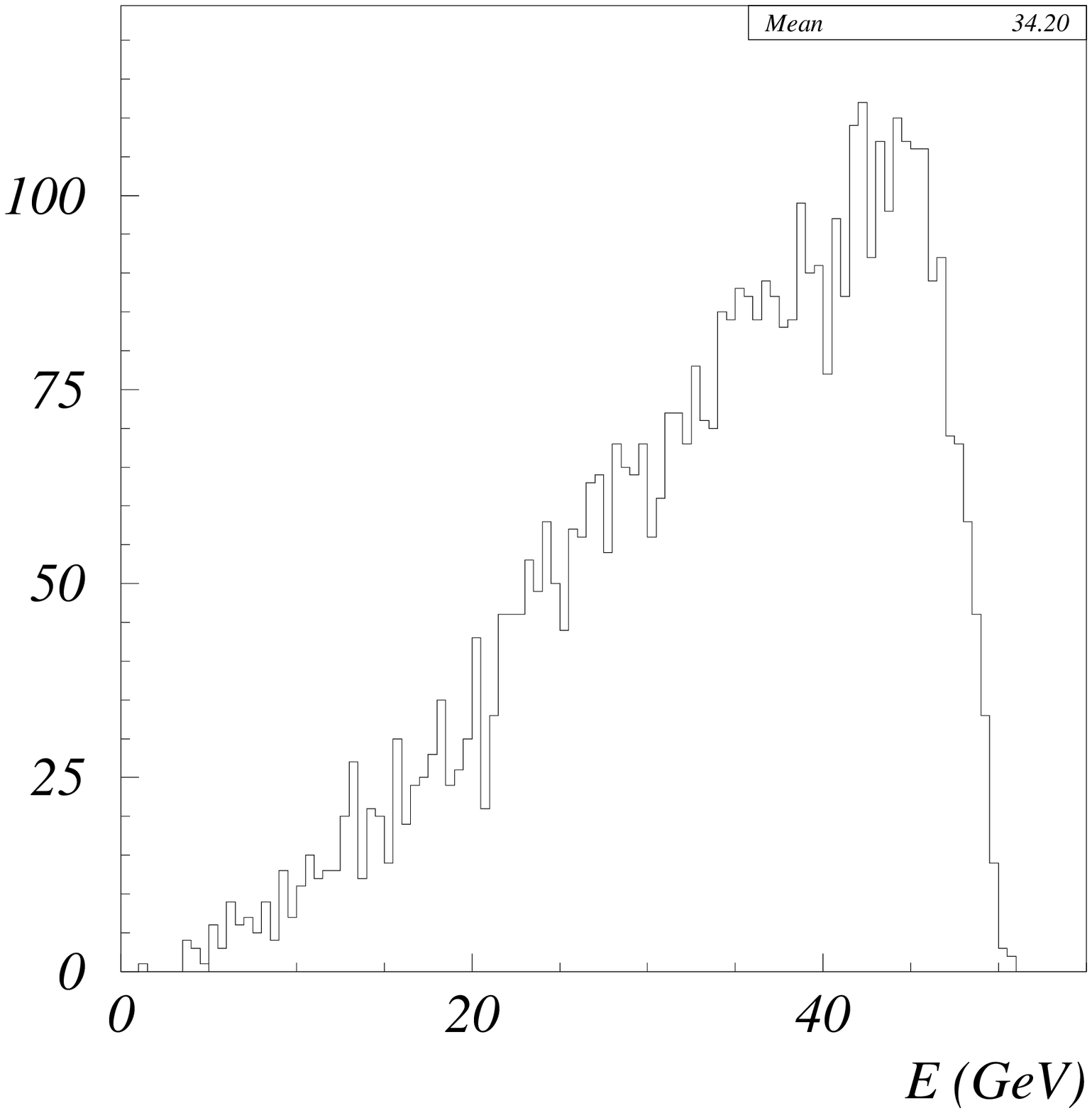}
}}
\end{center}
\caption{ Neutrino energy spectrum in the assumption of 1m diameter detector located 
1 km far from the neutrino source. }
\label{spectra}
\end{figure}

The target is made of $N$ emulsion bulks $1$~m diameter and $10$~cm 
thick each. By locating the detector 1 km  far from 
the neutrino source, we obtain the neutrino energy spectrum as shown 
in Fig.~\ref{spectra}.

The use of nuclear emulsions is limited by the interaction overlapping. 
A density of interactions of about $20~\mbox{cm}^3$ is affordable. Therefore, 
aiming at a statistics of $10^7~\bar{\nu}_{\mu}$ interactions we assume $N = 16$. 
For $1$~year data taking, a neutrino flux of $ \phi = 1.6 \times 10^{8}/16 (\nu/ s)$ 
through the detector would be required.  It corresponds to about $3.5\times 10^{14}$ 
decaying muons. Under these assumptions we obtain an overall density of 
interactions of about $16$ per $cm^3$, which is reasonable.

\subsection{Kinematical selection: deep-inelastic versus diffractive events}
\label{sec:kinema}

\begin{figure}[tb]
\begin{center}
\rotatebox{0}{\ \resizebox{0.7\textwidth}{!}{\
\includegraphics{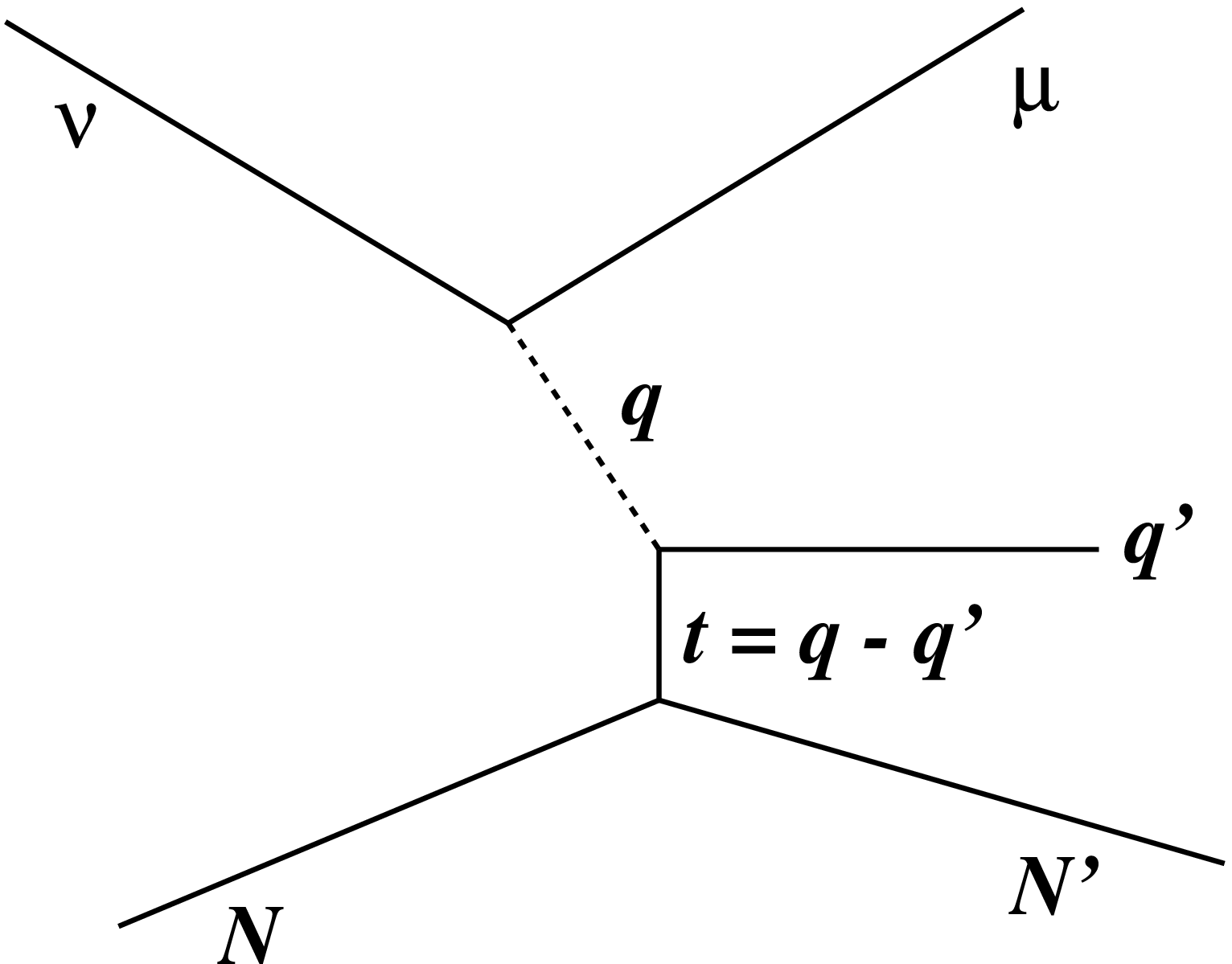}}}
\end{center}
\caption{The 4-momentum transfer $t$ in a neutrino CC 
interaction with charm production. $q^{\prime}$ is the charmed hadron
4-momentum. }
\label{fig:t}
\end{figure}

\begin{figure}[htbp]
\begin{center}
\rotatebox{0}{\ \resizebox{0.9\textwidth}{!}{\
\includegraphics{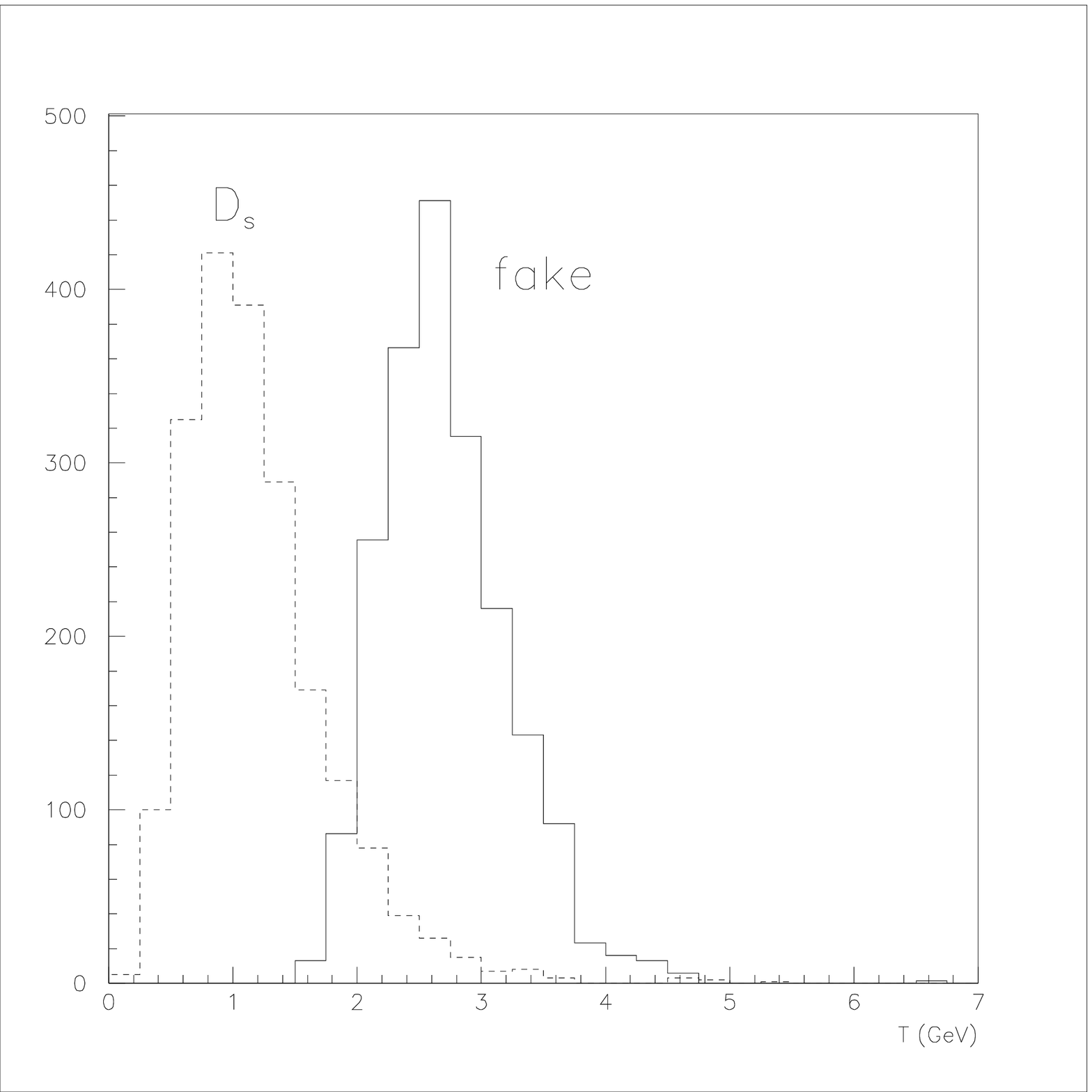}\includegraphics{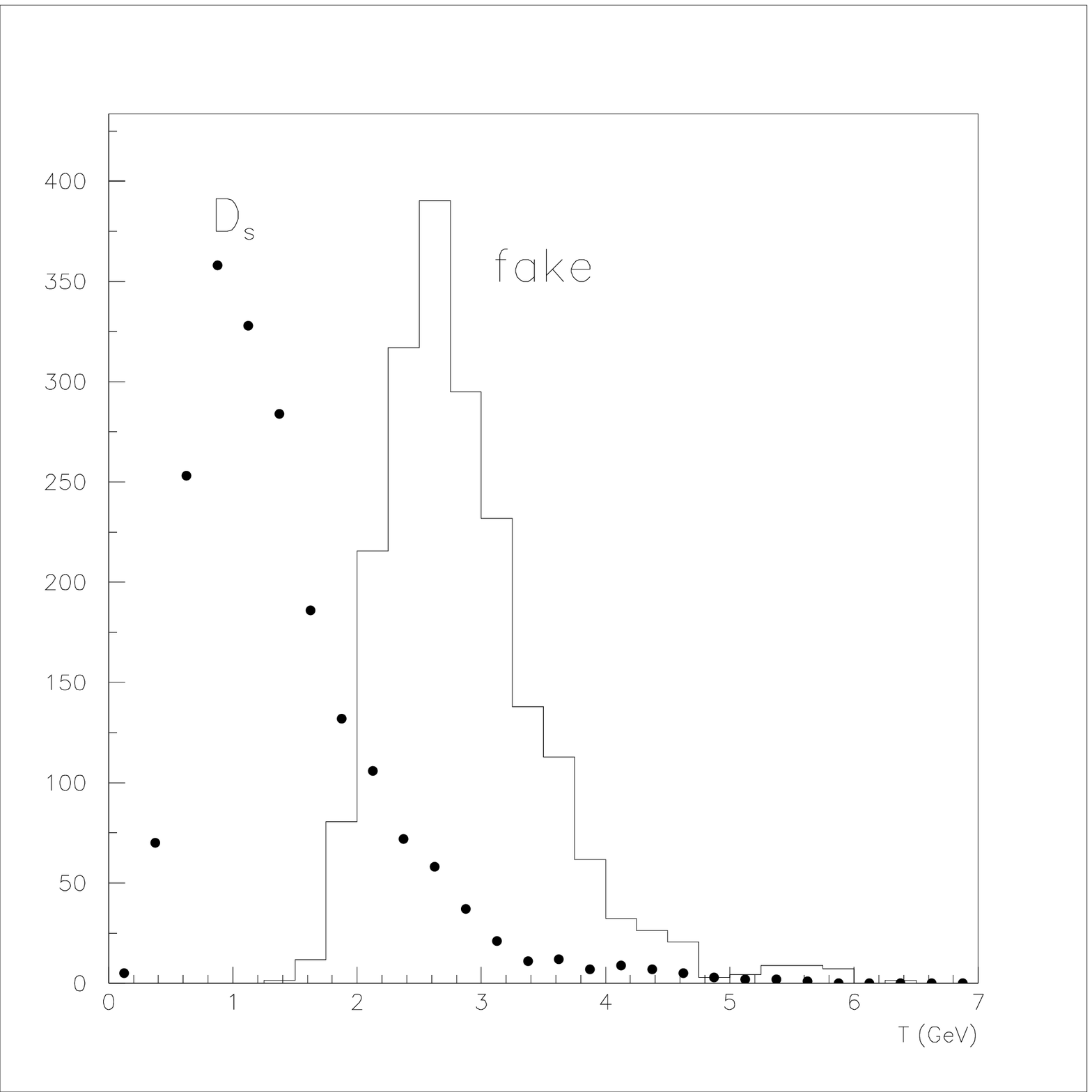}\includegraphics{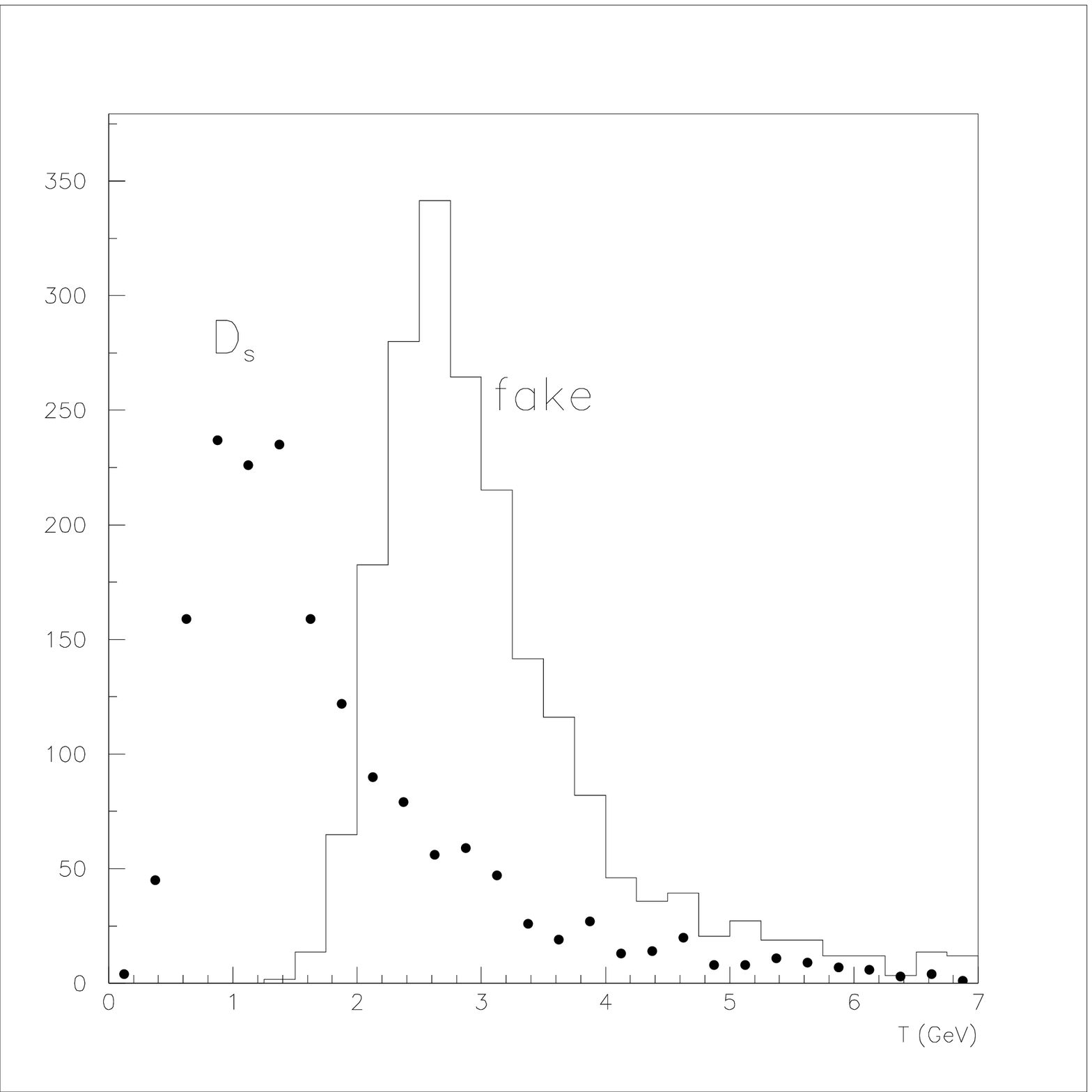}}}
\end{center}
\caption{$t$ distributions for both diffractive ($D_s$) and deep-inelastic 
({\it fake}) interactions with a diffractive topology. A resolution of 30\% 
(left plot), 50\% 
(middle one) and 100\% (right one) on the charmed hadron momentum measurement 
is assumed. }
\label{fig:tdis}
\end{figure}

Deep-inelastic charmed events are kinematically quite different with
respect to diffractive ones. In particular, the kinematical variable
$t$, defined in Fig.~\ref{fig:t}, allows a very high rejection
power against the background with a good efficiency of the signal. The
$t$ determination relies on the momentum determination of the charmed 
particle. In the E531 emulsion experiment~\cite{e531,e531_2}, a resolution in measuring the charmed hadron momentum better than $15\%$ was achieved  by using a likelihood technique. In the following we make the conservative assumption of momentum resolution in measuring the charmed hadron momentum of 
$30\%$, $50\%$ and
$100\%$. The reconstructed $t$ distributions for diffractive and
deep-inelastic events are shown in Fig.~\ref{fig:tdis}.

In order to select diffractive events we apply the cut $t<1.7$~GeV.
Cuts on the flight length ($> 10~\mu$m) and on the kink
angle ($\theta_{kink}> 15$~mrad) are also applied.
The efficiency for the signal and the correspondent background are
reported in Table~\ref{tab:kine}. From this table we can see that,
independently of the momentum resolution, the fraction of
deep-inelastic charm events with diffractive topology surviving the
kinematical cut  is $\varepsilon_{dis} \sim 0.4\%$ , while the signal
efficiency ranges from about $50$ to $80\%$. In the following we
assume a momentum resolution of $50\%$.

\begin{table}[htbp]
\begin{center}
{\small
\begin{tabular}{||c|c|c||}
\hline
$\Delta p/p$ & $\varepsilon_{dis} (\%)$ & $\varepsilon_{D_s} (\%)$ \\ \hline\hline
$30\%$ & $0.3\pm0.1$ & $83.3{\pm}0.8$ \\ \hline
$50\%$ & $0.4\pm0.2$ & $72.5{\pm}1.0$ \\ \hline
$100\%$ & $0.4\pm0.2$ & $51.8{\pm}1.1$ \\ \hline
\end{tabular}
}
\end{center}
\caption{Efficiency for the signal, $\varepsilon_{D_s}$, and for the 
background, $\varepsilon_{dis}$, assuming three different 
momentum resolutions.}
\label{tab:kine}
\end{table}

\subsection{Topology of neutrino induced diffractive charm events and background}

In the $D_s^{(*)}$ or $D^{(*)}$ diffractive production only a muon is
produced at the interaction point (primary vertex), besides the
charmed meson. Therefore, these events are characterised by a peculiar
topology: two charged tracks at the primary vertex, one of them being
a short lived particle.

Note that the excited states produced undergo the following decays
without leaving an observable track:

\begin{eqnarray*}
  D^{*-}_s\rightarrow D^{-}_s\gamma & BR = (94.2\pm2.5)\%~\cite{pdg2000}\\
  D^{*-}_s\rightarrow D^{-}_s\pi^0 & BR = (5.8\pm2.5)\%~\cite{pdg2000}\\
  D^{*-}\rightarrow D^{0}\pi^- & BR = (67.7\pm0.5)\%~\cite{pdg2000}\\
  D^{*-}\rightarrow D^{-}\pi^0 & BR = (30.7\pm0.5)\%~\cite{pdg2000}\\
  D^{*-}\rightarrow D^{-}\gamma & BR = (1.6\pm0.4)\%~\cite{pdg2000}. 
\end{eqnarray*}

A source of irreducible background is the diffractive
production of $D^{(*)-}$. 
Its contamination relatively to the signal is given by 
\[ 
\varepsilon_{D^{(*)}} = \left| \dfrac{V_{cd}}{V_{cs}} \right|^2 
\times [ \eta_{D^{-}} + \eta_{D^{*-}}\times 
BR(D^{*-} \rightarrow D^-) ]
\]
where $BR(D^{*-} \rightarrow D^-) = 0.323\pm0.006$~\cite{pdg2000},
$\eta_{D^{-}}$ and $\eta_{D^{*-}}$ are the fractions of diffractively
produced $D^{-}$ and $D^{*-}$ respectively. We assume that these
fractions are the same for $D^{-}$ as for $D_s^{-}$
($\eta_{D^{-}}=\eta_{D^{*-}}=0.5$). The latter can be extracted by the
NuTeV results: $\sigma(\nu_\mu N\rightarrow\mu^- D_s N) =
(1.4\pm0.3)$fb/nucleon and $\sigma(\nu_\mu N\rightarrow\mu^- D_s^* N)
= (1.6\pm0.4)$fb/nucleon~\cite{dsprod_nutev}. Finally we get
$\varepsilon_{D^{*}} = (3.3\pm0.8)\%$.

\begin{figure}[htbp]
\begin{center}
\rotatebox{0}{\ \resizebox{0.7\textwidth}{!}{\ \includegraphics{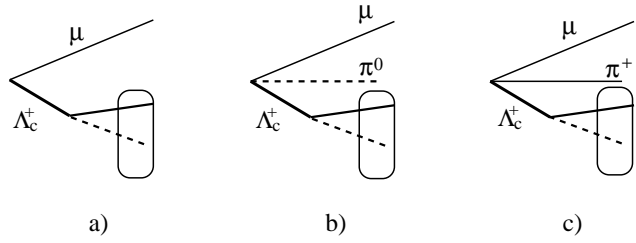} }}
\end{center}
\caption{Topology of the quasi-elastic charm neutrino induced events in the
case of the reaction a) $\nu_\mu n\rightarrow\mu^-\Lambda_c^+$, b) $\nu_\mu
n\rightarrow\mu^-\Sigma_c^+(\Sigma_c^{*+})$ and c) $\nu_\mu
p\rightarrow\mu^-\Sigma_c^{++}(\Sigma_c^{*++})$. The particles inside the
box represent the $\Lambda_c^+$ decay products.}
\label{fi:topo}
\end{figure}

Neutrino induced quasi-elastic charm events are characterised by the
topologies shown in Fig.~\ref{fi:topo} (see Ref.~\cite{lambda}).
Therefore, they are similar to the diffractive ones, but with a
cross-section twice as large. Since anti-neutrinos cannot induce
quasi-elastic charm production while diffractive production is the 
same for both $\nu$ and $\bar{\nu}$, our interest is focused on 
$\bar{\nu}$ induced events. Therefore we make the assumption that
all the events with the above topology are due to $D_s^{(*)}$ diffractive
production. 

The charm production in $\bar{\nu}$ deep-inelastic interactions and
the event topology have been studied by using the HERWIG event
generator~\cite{herwig}, an event generator based on
JETSET~\cite{jetset} and LEPTO~\cite{lepto} and the energy dependence
of the charmed fractions reported in Ref.~\cite{bolton}\footnote{We
assume that the charmed fractions are equal for both $\nu$ and
$\bar{\nu}$ as implemented in the event generators.}.  The average
charmed fractions, convoluted with the anti-neutrino
spectrum~\cite{bigi}, are $F_{\bar{D}^0} = 61\%$, $F_{{D}^-} = 26\%$,
$F_{{D}^-_s} = 7.3\%$ and $F_{{\Lambda}^-_c} = 5.7\%$.  The kinematics
signal has been modeled according to
Refs.~\cite{teods1,teods2,teods3,teods4,teods5} by using an event
generator developed within the CHORUS Collaboration~\cite{chorus_ds}.

The contamination of the diffractive sample from deep-inelastic events 
can be written as  

\begin{equation}
\varepsilon_{fake} = \frac{\sigma(\bar{\nu}_\mu N\rightarrow\mu^+ C X)}{
\sigma(\bar{\nu}_\mu N\rightarrow\mu^+X)}{\times}\frac{1}{\bar{\mathcal{R}}}{\times}
 (F_{D^-}+F_{\Lambda_c^-}) \times f_{fake}{\times}\varepsilon_{dis}
\label{eq:eprio}
\end{equation}

where 
$$\bar{\mathcal{R}}\equiv\frac{\sigma(\bar{\nu}_\mu
  N\rightarrow\mu^+D_s^{(*)-} N)}{\sigma(\bar{\nu}_\mu
  N\rightarrow\mu^+X)}\,.$$

We take the charm production in $\bar{\nu}$ interaction to be
$3\%$~\cite{conrad}. By using the charmed fractions given above,
$(F_{D^-}+F_{\Lambda_c^-})=31.7\%$. The factor $f_{fake} = (6.0\pm
0.1)\%$ denotes the fraction of deep-inelastic charmed events faking a
diffractive topology. $\varepsilon_{dis}$ gives the efficiency of
kinematical cuts as explained in Section~\ref{sec:kinema}.

As discussed in Section~\ref{diff_pro}, the experimental determination
of $\bar{\mathcal{R}}$ has an accuracy of about $15\%$ which gives
$\varepsilon_{fake} = (0.037\pm0.009)\%$. The $\varepsilon_{fake}$
value and its error are reported in the third column of
Table~\ref{tab:erro2} as a function of the error on
$\bar{\mathcal{R}}$.

From the numbers given above it turns out that the little knowledge we
have about the diffractive charm production cross-section plays a role
only in the evaluation of the deep-inelastic contamination, namely a
term of the systematic error. Even if the ratio $\bar{\mathcal{R}}$
had an uncertainty of $100\%$, at $3\sigma$ the relative systematic
error on the branching ratios would be $\lesssim 0.16\%$ (see
Table~\ref{tab:erro2}), so that the $\varepsilon_{D^{(*)}}$
contribution is dominant. Therefore, the overall systematic
uncertainty does not depend at all on the $\bar{\mathcal{R}}$ accuracy
($\varepsilon_{sys} = (3.3\pm0.8)\%$).

\begin{table}[tbp]
\begin{center}
{\small
\begin{tabular}{||c|c|c||}
\hline $\Delta\bar{\mathcal{R}}/\bar{\mathcal{R}}$ &
$\sigma_{fake}/\varepsilon_{fake}$ &
$\varepsilon_{fake} (\%)$ \\ 
\hline\hline 
$15\%$ & $22\%$ & $0.037{\pm}0.009$ \\ 
\hline 
$30\%$ & $34\%$ & $0.04{\pm}0.01$ \\
\hline 
$50\%$ & $53\%$ & $0.04{\pm}0.02$ \\ 
\hline 
$100\%$ & $101\%$ & $0.04{\pm}0.04$ \\ 
\hline
\end{tabular}
}
\end{center}
\caption{The relative and absolute error on $\varepsilon_{fake}$ as
  a function of the relative error on $\bar{\mathcal{R}}$.}
\label{tab:erro2}
\end{table}

\subsection{Description of the method}
\label{metho2}

An almost pure sample of $D_s^-$ from diffractive events, with a small
contamination of $D^-$ and $\Lambda_c^-$ produced in deep-inelastic
events and in diffractive $D^{(*)-}$ production, can be built by using
diffractive $D_s^{(*)}$ production from anti-neutrinos. The
normalisation to determine the $D_s$ absolute branching ratios is
given by the number of events with a vertex topology consistent with
one $\mu$ plus a short lived particle. No model dependent information
is used to define the normalisation.

It is also worth noting that, in particular, the contamination of
 $D^-$ and $\Lambda_c^-$ events does not affect the
 $D_s\rightarrow\tau$ channel. Indeed, such events would present a
 unique topology with two subsequent kinks. An event with a double
 kink has been recently observed in CHORUS (see
 Ref.~\cite{chorus_ds}).

\subsection{Measurement accuracy at a neutrino factory}

At present there are no experiments with both an adequate spatial 
resolution to fully exploit the diffractive topology and a sufficient 
anti-neutrino induced CC event statistics. Therefore, the 
method proposed in this paper could only be exploited with the above-mentioned 
detector exposed at a neutrino factory. 

Let us assume to collect $10^7$ $\bar{\nu}$CC events into an emulsion
target and to have a detection efficiency of about $73\%$ for the
$D_s$ decays (see Table~\ref{tab:kine}).  By assuming a vertex
location efficiency of about $50\%$\footnote{This efficiency accounts
for the electronic detector reconstruction and the automatic location
of the event vertex inside the emulsions.} and assuming a $\bar{\nu}$
diffractive production rate of $6.2\times10^{-3}$/CC, we expect to
detect a number of $D_s$ equal to $N_{D_s} = 10^7\times
6.2\times10^{-3}\times0.73 \times 0.5\simeq 2.3\times10^4$.

The expected accuracy on the determination
of the $D_s$ branching ratios is shown in Table~\ref{tab:brds} for a few 
channels, together with the current status. To compute the expected number of
events in each decay channel we have used the central values (shown in
Table~\ref{tab:brds} together with their errors) given by the
Particle Data Group~\cite{pdg2000}.

\begin{table}[tbp]
\begin{center}
{\small
    \begin{tabular}{||c|c|c||}
      \hline
      Channel & PDG BR~\cite{pdg2000} & This method \\
      \hline
      \hline
      $D_s\rightarrow \mu\nu$ & $(4.6 \pm 1.9)\times10^{-3}$ & 
$({\pm}0.55{\pm}0.15)\times10^{-3}$ \\
      \hline
      $D_s\rightarrow \tau\nu$ & $(7 \pm 4)\%$ & 
$({\pm}0.17{\pm}0.23)\%$  \\
      \hline
      $D_s\rightarrow \phi l\nu$ & $(2.0 \pm 0.5)\%$ & 
$({\pm}0.08{\pm}0.07)\%$ \\
      \hline
      \hline
    \end {tabular}
}
\end{center}
\caption{Statistical and systematic accuracy achievable in the
  determination of the $D_s$ absolute branching ratios, assuming a
  collected statistics of $10^7~\bar{\nu}_\mu$ CC events. 
The central values are  taken from Ref. \cite{pdg2000}.}
\label{tab:brds}
\end{table}

By using the relation given in Section~\ref{leptodec} and the measured
branching ratios given in Table~\ref{tab:brds}, the decay constant
$f_{D_s}$ can be extracted. If we collect $10^7$ $\bar{\nu}_{\mu}$ CC
interactions we get $f_{D_s} = 288\pm4\mid _{stat}\pm5\mid
_{sys}~$MeV where the central value is taken from
Ref.~\cite{pdg2000}.

%

\section{Conclusions}
$D_s$ branching ratios are affected by large uncertainties, mainly due
to the difficulty of defining a pure $D_s$ starting sample. These
uncertainties translate into the $f_{D_s}$ determination which is in
turn affected by large errors.  A method for the evaluation of the
$D_s$ branching ratios and of the $f_{D_s}$ decay constant with a
systematic accuracy better than $5\%$ has been presented.

The idea is to build an almost pure sample of $D_s$'s by means of the
anti-neutrino induced diffractive $D_s^{(*)}$ production. The vertex
topology of these events is extremely simple: there is only a muon and
a short lived charmed hadron, the $D_s$. This peculiarity makes the
contamination of $D^{-}$ and $\Lambda_c^{-}$ from anti-neutrino
deep-inelastic interactions negligible. The diffractive $D^{(*)}$
production yields a contamination of about $3\%$ which is the dominant
systematic uncertainty.

In order to make the statistical error at the same level as the
systematic one, a copious ($\mathcal{O}(10^{7})$) number of
anti-neutrino interactions is needed.

At present there are no experiments with both an adequate spatial 
resolution to fully exploit the diffractive topology and a sufficient 
anti-neutrino induced CC event statistics. Therefore, the measurement could 
only be performed at a neutrino factory. 

%

\section*{Acknowledgements}
We would like to thank S. Anthony for the careful reading of the manuscript.
%
%


\end{document}